\documentclass[twocolumn,showpacs,prl]{revtex4}

\usepackage{amsmath}

\begin{document}

\title{Sampling from naturally truncated power laws: The matchmaking paradox}
\author{I.M. Sokolov}
\affiliation{Institut f\"ur Physik, Humboldt-Universit\"at zu Berlin, Newtonstr. 15,
D-12489 Berlin, Germany}
\author{I. Eliazar}
\affiliation{Department of Technological Management, Holon Institute of Technology,
P.O.B. 305, IL-58102 Holon, Israel}

\begin{abstract}
Consider a network of $M\gg 1$ nodes connected by $N\gg 1$ links, in which
the distribution of the number of links per node follows a power law $%
P(n)\simeq n^{-1-\alpha }$ with exponent $0<\alpha <1$. The power law is
naturally truncated due to the fact that $N$ is finite. A subset of $m\ll M$
nodes is sampled arbitrarily, yielding the sample mean $\eta $: The average
number of links per node, within the sampled subset. We explore the
statistics of the sample mean $\eta $ and show that its fluctuations around
the population mean $\nu =N/M$ are extremely broad and strongly skewed --
yielding typical values which are systematically and significantly smaller
than the population mean $\nu $. Applying these results to the case
of bipartite networks, we show that the sample means of the two parts of
these networks generally differ -- the fact we call \textquotedblleft matchmaking paradox\textquotedblright 
in the title.
\end{abstract}

\maketitle

In this letter we address the problem of sampling random networks with
naturally truncated power-law distributions of the number of links. Consider
a network consisting of $M\gg 1$ nodes connected by $N\gg 1$
links.\emph{\ }Imagine that in a very large population ($M\rightarrow \infty $, 
$N\rightarrow \infty $, and $N/M\rightarrow \nu=const$) 
the distribution of the number of links tends to a power law 
\begin{equation*}
P(n)\simeq n^{-1-\alpha }
\end{equation*}
for $n$ large enough (with a normalization coefficient which might
depend on the actual network size). In a finite population, the finite mean 
$\nu =N/M$ implies that the distribution of $n$ is truncated at some value:
This is what we term a naturally truncated power-law distribution. Natural
truncation has to be distinguished from the finite size effects in growing
networks (see e.g. \cite{Waclaw} and references therein).

Imagine moreover that -- like is always done in statistical investigations
-- a random sample of size $m\leq M$ nodes is drawn from the overall
population of $M$ nodes. If the corresponding power-law exponent is in the
range $0<\alpha <1$, then the mathematical expectation $\sum_{n=0}^{\infty
}nP(n)$ diverges -- and hence fails to coincide with the population mean 
$\nu $. In such a situation we inquire the distribution of the sample mean $%
\eta =\frac{1}{m}\sum_{i=1}^{m}n_{i}$, where $n_{i}$ denotes the number of
links connected to the $i^{\text{th}}$ node of the sample. In
particular, it is of interest to know whether the sample mean $\eta $ is
typically larger or smaller than the population mean $\nu $, and how do its
statistics change as the sample size $m$ is increased.

The aforementioned problem is related to the \textquotedblleft L\'{e}vy
matchmaking\textquotedblright\ problem. Imagine two sets of $M\gg 1$ nodes
(the red and the blue nodes, or boys and girls). The nodes of the two sets
are connected by $N\gg 1$ links having a red node on one side and a blue one
the other side. Although the number of the links is the same when seen from
the red and from the blue side, the distributions of the number of links
attached to a red and to a blue nodes differ. In a very large population ($%
M\rightarrow \infty $ , $N\rightarrow \infty $, and $N/M\rightarrow \nu
=const$) they would follow 
\begin{equation*}
P_{\mathrm{red}}(n)\simeq n^{-1-\alpha _{1}};\qquad P_{\mathrm{blue}%
}(n)\simeq n^{-1-\alpha _{2}}
\end{equation*}%
for $n$ large enough (with, in general, exponents $\alpha _{1}\neq \alpha
_{2}$). In such a situation -- when sampling from the red and blue
populations -- how different can the red and are the sample means be?

The motivation for the L\'{e}vy matchmaking problem is as follows. In 
the mid eighties several research groups were conducting
investigations on the distribution of the number of sexual partners in
different human populations promoted by the necessity to point out the risk
groups in the AIDS epidemics. A \textquotedblleft Nature\textquotedblright\
editorial by Maddox's contained the statement \cite{Maddox}:
\textquotedblleft \textit{The figures so far show that the average number of
heterosexual partners of men in the course of a lifetime is 11.0 and of
women 2.9}\textquotedblright . In response to Maddox' editorial, 
Gurman published a note explaining the nonsense of having different means
in the two populations connected by well-defined one-to-one links \cite%
{Gurman}: \textquotedblleft \textit{A heterosexual union is analogous to a
heteronuclear chemical bond, and the total number must be the same if viewed
from the male or female end}\textquotedblright .

This situation is more profound than it seemed to be. The empirical
distribution of the number of partners is long-tailed \cite{Morris}, follows
a power-law, and its mathematical expectation may diverge. Thus for
exponents in the range $0<\alpha <1$ the sample means depend systematically
on the sample size $m$ and therefore \textit{have to differ} for small
samples in order to match each other for the population as a whole. This
point is what we refer to as the \textquotedblleft \textit{matchmaking
paradox}\textquotedblright\ in the title of this Letter. For exponents in
the range $\alpha >1$ this is no more the case, and the sample means have to
match. Up to our best knowledge, this exponent-dependency aspect of the
problem was never considered in detail (probably due to the lack,
at that time, of an adequate mathematical toolbox). Moreover, the problem
has much in common with other situations of weak ergodicity breaking.
Indeed, sample means that \textquotedblleft normally\textquotedblright\
should be the same, actually differ since one of them never reaches a sharp
value but shows universal fluctuations \cite{Barkai, Lubelski,He}.

Later investigations \cite{Liljeros1, Schneeberger} have shown that
power-laws in a heterosexual population have exponents in the range $\alpha
>1$, implying that the reason for sample-mean deviations should be looked
for elsewhere (see e.g. Ref. \cite{Wiederman}). Nonetheless, both the
problems of sampling and matchmaking are of considerable interest --
especially taking into account the overall importance of the sampling
procedures in networks \cite{Stumpf}, as well as the fact that the
distribution of the number of contacts in homosexual males follows a
power-law with exponent $\alpha \approx 0.6$, Ref.\cite{Schneeberger}.

The main issues explored in this research are the following: What is a
distribution of a sample mean $\eta $ calculated for a sample of size $m\gg 1
$? And how does the sample mean $\eta $ relate to the population mean $\nu $
? These issues are intimately connected to the statistics of 
\textit{L\'{e}vy random probabilities}, studied in Ref. \cite{Eliazar} --
but have several unique aspects which are worth a separate and detailed
investigation.

We follow Gurman's setup with a static, finite, bipartite
population. To begin with, we establish a model yielding naturally truncated
power-law distributions (of the links). Consider a large population
consisting of $2M\gg 1$ nodes -- $M$ \textquotedblleft
red\textquotedblright\ and $M$ \textquotedblleft blue\textquotedblright\ --
and $N\gg 1$ links connecting the red and blue nodes. Each node has
an \textquotedblleft attractiveness\textquotedblright\ level: Each
red node $i$ (blue node $j$) has an attractiveness level 
$f_{i}$ ($g_{j}$) chosen at random from a one-sided L\'{e}vy 
distribution with exponent $\alpha _{1}$ ( $\alpha _{2}$). 
Each link connects -- on each red/blue side -- to a single node, the
probability of connecting being proportional to the attractiveness levels.
Hence, the probabilities $\phi _{i}$ and $\gamma _{j}$ that the ends of a
given link are connected to the red node $i$ and to the blue node $j$ are
given by 
\begin{equation*}
\phi _{i}=\frac{f_{i}}{\sum_{k=1}^{M}f_{k}},\qquad \gamma _{j}=\frac{\mathbf{
g}_{j}}{\sum_{k=1}^{M}g_{k}}.
\end{equation*}

Let us first concentrate on the red side of the network. As a statistical
sample we chose at random a set of $m<M$ of the red nodes. The probability
that a given link is connected to one of the sample nodes is given by 
\begin{equation*}
p_{i}=\frac{\sum_{i=1}^{m}f_{i}}{\sum_{j=1}^{M}f_{j}}=\frac{
\sum_{i=1}^{m}f_{i}}{\sum_{j=1}^{m}f_{j}+\sum_{j=m+1}^{M}f_{j}}=\frac{1}{
1+Y/X},
\end{equation*}
where $X$ and $Y$ are the independent one-sided L\'{e}vy variables with
exponent $\alpha =\alpha _{1}$, and with scaling parameters $m^{1/\alpha }$
and $(M-m)^{1/\alpha }$. The value of $p_{i}$ -- the L\'{e}vy random
probability -- is thus a random variable which coincides in distribution
with 
\begin{equation*}
z=\frac{1}{1+\left( M/m-1\right) ^{1/\alpha }R},
\end{equation*}
where $R$ is quotient of two independent one-sided L\'{e}vy variables with
exponent $\alpha =\alpha _{1}$. Henceforth, we set the shorthand notation 
$x=\left( M/m-1\right) ^{1/\alpha }$. Note that the random variable $z$
admits values in the unit interval $\left( 0,1\right) $. Moreover, we note
that even if the distributions of the attractiveness levels $f_{i}$ deviate
from the one-sided L\'{e}vy -- but yet possess power-law
asymptotics with exponent $\alpha $ -- then the distribution of $z$ for $%
m,M\gg 1$ is universal (in the sense of the corresponding limit theorem).
Hence, our analysis does not depend on the precise form of distributions of
the attractiveness levels $f_{i}$ . We further note that the introduction of
the attractiveness levels was only a convenient intermediate step, and that
the discussion to follow holds for any kind of naturally truncated power-law
distributions with exponents in the range $0<\alpha <1$.

The probability density function (pdf) of the quotient $R$ is known \cite%
{Eliazar}: Its Laplace transform is a Mittag-Leffler function $\mathcal{L}%
\{p_{R}(R)\}=E_{\alpha }(-u^{\alpha })$ with $u$ denoting the Laplace
variable. And, the asymptotic behavior of $p_{R}$ for $R$ large and small is
obtained via Tauberian theorems from the asymptotics of the Mittag-Leffler
function. Thus, for $R$ large we have 
\begin{equation}
p_{R}(R)=\frac{1}{\Gamma ^{2}(\alpha )}R^{-1-\alpha },  \label{pofR}
\end{equation}
where $\Gamma (\cdot )$ is a Gamma function.

Let $h=\sum_{i=1}^{m}n_{i}$ denote the number of \textquotedblleft
hits\textquotedblright\ in the sample. Given the value $z$ of the
probability of connecting to one of the sample nodes, the probability that $h
$ of $N$ links \textquotedblleft hit\textquotedblright\ the sample is given
by the conditional binomial distribution 
\begin{equation*}
p(h|z)=\frac{N!}{h!(N-h)!}z^{n}(1-z)^{N-h}.
\end{equation*}
Hence, the unconditional probability distribution of $h$ is given by 
\begin{equation*}
p_{h}(h)=\int_{0}^{\infty }\frac{N!}{h!(N-h)!}z^{n}(1-z)^{N-h}p_{z}(z)dz.
\end{equation*}
For $N\gg 1$ the binomial distribution is actually extremely narrow: Its
standard deviation is much smaller than its mean, so that 
$[N!/h!(N-h)!]z^{h}(1-z)^{N-h}\approx \delta (h-Nz)$. Thus we can take $h=Nz$
; the distribution of $h$ follows from those of the L\'{e}vy random
probability $z$ by change of variables. The distribution of the sample mean 
$\eta =h/m=Nz/m$, in turn, is given by 
\begin{equation*}
p_{\eta }(\eta )\approx \frac{m}{N}p_{z}\left( \eta \frac{m}{N}\right) .
\end{equation*}
This fact can be proved by explicit calculation of the generating function
of the distribution $p_{h}\left( \cdot \right) $ -- evaluating it in the
range $1\ll h\ll N$ via Tauberian theorems.

Note that for $M\rightarrow \infty $ and $m\ll M$ $p_{z}(z)$ practically
follows the distribution of $M^{-1/\alpha }R$, and is a power-law. Taking 
$m=1$ we arrive at the (continuous approximation for the) distribution of the 
number of links per node. The power law spreads over the domain of 
$1\ll h\ll N$ and is truncated for $h>N$, as it is evident from the fact that 
$p_{z}(z)$ vanishes for $z>1$. The sample mean $\eta $ is therefore a random
variable, and the properties of its distribution are discussed
below.

The mathematical expectation $\left\langle \eta \right\rangle $ of
the sample mean $\eta $ is \textit{equal} to the population mean 
$\nu $. Indeed, 
\begin{equation*}
\left\langle \eta \right\rangle =\left\langle \frac{\mathbf{h}}{m}
\right\rangle =\frac{N}{m}\left\langle z\right\rangle ,
\end{equation*}
and 
\begin{equation}
\left\langle z\right\rangle =\int_{0}^{\infty
}z(R)p_{R}(R)dR=\int_{0}^{\infty }\frac{1}{1+xR}p_{R}(R)dR.  \label{EqR}
\end{equation}
Noting that $1/(1+xR)=x^{-1}(1/x+R)^{-1}$ and substituting
the integral representation  
\begin{equation}
\frac{1}{1/x+R}=\int_{0}^{\infty }e^{-u/x}e^{-uR}du  \label{repres}
\end{equation}
into Eq.(\ref{EqR}) -- while interchanging the order of
integration -- yields:
\begin{eqnarray}
\left\langle z\right\rangle &=& \frac{1}{x}\int_{0}^{\infty
}due^{-u/x}\int_{0}^{\infty }dRe^{-uR}p_{R}(R) \nonumber \\
&=& \frac{1}{x}\int_{0}^{\infty } du e^{-u/x} E_{\alpha }(-u^{\alpha }).
\label{EqZ}
\end{eqnarray}
The right-hand-side of Eq. (\ref{EqZ}) is the Laplace transform of
this Mittag-Leffler function. This Laplace transform is known to be given by 
$\mathcal{L}\{E_{\alpha }(-u^{\alpha })\}=s^{\alpha -1}/(s^{\alpha }+1)$, 
and hence setting $s=1/x$ we arrive at
\begin{equation*}
\left\langle z\right\rangle =\frac{1}{x^{\alpha }+1}.
\end{equation*}
Finally, recalling that $x=(M/m-1)^{1/\alpha }$ we obtain that $\left\langle
z\right\rangle =m/M$ and 
\begin{equation*}
\left\langle \eta \right\rangle =\frac{N}{m}\frac{m}{M}=\frac{N}{M}=\nu .
\end{equation*}

The distribution of the sample mean $\eta $, however, is extremely broad --
as seen from its variance. To calculate the variance we note that 
$\left\langle \eta ^{2}\right\rangle =(N^{2}/m^{2})\left\langle
z^{2}\right\rangle $ and 
\begin{equation*}
\left\langle z^{2}\right\rangle =\int_{0}^{\infty
}z(R)p_{R}(R)dR=\int_{0}^{\infty }\frac{1}{(1+xR)^{2}}p_{R}(R)dR.
\end{equation*}
Using the fact that $(1+xR)^{2}=\frac{d}{dx}(1/x+R)^{-1}$ and the integral
representation given by Eq.(\ref{repres}) we get: 
\begin{equation*}
\left\langle z^{2}\right\rangle =\frac{d}{dx}\frac{x}{1+x^{\alpha }}=\frac{
m^{2}}{M^{2}}\left[ 1+(1-\alpha )\left( \frac{M}{m}-1\right) \right] .
\end{equation*}
From this we obtain that the variance of $\eta $ is given by: 
\begin{equation*}
\sigma ^{2}=\left\langle \eta ^{2}\right\rangle -\left\langle \eta
\right\rangle ^{2}=(1-\alpha )\frac{N^{2}}{M^{2}}\left( \frac{M}{m}-1\right)
\simeq (1-\alpha )\nu ^{2}\frac{M}{m}.
\end{equation*}
Hence, the standard deviation $\sigma $ of $\eta $ is of the order
of magnitude $\sqrt{M/m}\gg 1$ -- i.e., far larger than its mean 
$\left\langle \eta \right\rangle $. Therefore, it is highly improbable to
obtain an accurate estimate of the population mean $\nu $ from a
sample with size \thinspace $m$ much smaller than the population size $M$.

Not only is the distribution of $\eta $ extremely broad -- it is also
extremely skewed. As we now proceed to show, the median of $\eta $ lays far
below its mathematical expectation $\left\langle \eta \right\rangle =\nu $.
And, finding values of $\eta $ which are larger than its
mathematical expectation $\left\langle \eta \right\rangle =\nu $ is highly
improbable. Hence, a typical result of a statistical measurement of $\eta $
will be much smaller than the population mean $\nu $.

Since $z$ and therefore $\eta $ are monotonous functions of $R$, their
medians follow from the median of $R$. The random variable $R$ is a quotient
of two identically distributed random variables -- hence the distribution of 
$R$ is the same as the distribution of $1/R$. The random variable $\ln (R)$
is therefore symmetric and, consequently, its median is zero -- implying, in
turn, that the median of $R$ is unity: $R_{1/2}=1$. Substituting the median 
$R_{1/2}=1$ into the expressions for $z$ and $\eta =(N/M)z$ we obtain that
the median $\eta _{1/2}$ of the sample mean $\eta $ is given by:
\begin{equation}
\eta _{1/2}=\frac{N}{m}\frac{1}{1+x}\simeq \nu \left( \frac{m}{M}\right) ^
{\frac{1}{\alpha }-1}  \label{median}
\end{equation}
(equation (\ref{median}) holding for all $m\ll M$). Clearly, the median 
$\eta _{1/2}$ is much smaller than the population mean $\nu $.

Let us turn now to calculate the probability $P_{+}$ that that the sample
mean $\eta $ be greater than the population mean $\nu $ -- i.e., the
probability of the event $\left\{ \eta >\nu \right\} $. Using the asymptotic
expression for $p_{R}(R)$ gives 
\begin{equation*}
P_{+}=\int_{m/M}^{\infty }p_{z}(z)dz=\int_{m/M}^{\infty
}p_{z}(z)dz=\int_{R_{0}}^{\infty }p_{R}(R)dR
\end{equation*}
with $R_{0}=\left( M/m-1\right) ^{1-1/\alpha }$. Further using Eq.(\ref{pofR}), 
we obtain that 
\begin{equation}
P_{+}\simeq \frac{1}{\alpha \Gamma ^{2}(\alpha )}\left( \frac{M}{m}-1\right)
^{\alpha -1},  \label{pplus}
\end{equation}
(equation (\ref{pplus}) holding for all $m\ll M$). Clearly, the probability 
$P_{+}$ is very small.

Thus, in a L\'{e}vy matchmaking problem, the sample means in different
subpopulations not only fluctuate strongly, but also display a systematic
difference. For the same sample size, the subpopulation with smaller $\alpha 
$ -- i.e., the one with a broader distribution -- will typically show a
smaller sample mean. The discussion above also gives a possibility to 
roughly estimate the unknown population mean $\nu $ from the
typically smaller sample mean $\eta $. Such an extrapolation is given by 
Eq.(\ref{median}) (or by Eq.(\ref{example}) -- in the special L\'{e}vy-Smirnov
case).

It is instructive to consider an analytically solvable example -- the L\'{e}%
vy-Smirnov case, corresponding to the exponent value $\alpha =0.5$. This
example is of special interest due to the fact that the exponent $\alpha
=0.5 $ is not too far from the exponent $\alpha \approx 0.6$ obtained from
the distribution of the number of partners in the population of homosexual
males. The L\'{e}vy-Smirnov pdf of the attractiveness levels is given by%
\begin{equation*}
p(f)=\frac{1}{2\sqrt{\pi }f^{3/2}}\exp \left( -\frac{1}{4f}\right) ,
\end{equation*}
for which 
\begin{equation*}
p_{z}(z)=\frac{1}{\pi }\sqrt{\frac{x}{z(1-z)}}\frac{1}{1+(x-1)z}.
\end{equation*}
The quantiles of the corresponding distributions can be calculated
explicitly -- implying, in turn, that with probability $0.5$ the sample mean 
$\eta $ lays within the interval 
\begin{equation}
(\sqrt{2}-1)^{2}\nu \frac{m}{M}<\eta <(\sqrt{2}+1)^{2}\nu \frac{m}{M}.
\label{example}
\end{equation}
Namely, the sample mean $\eta $ is typically considerably smaller than the
population mean $\nu $. Only as $m\rightarrow M$ does the median $\eta _{1/2}$
converge to the population mean $\nu $. On the other hand, the distribution
over samples is very skewed, and the probability that the sample mean $\eta $
be greater than the population mean $\nu $ is given by $P_{+}\simeq (2/\pi )%
\sqrt{m/M}$. Namely, $P_{+}$ is very small for sample sizes $m$ which are
considerably smaller than the population size $M$.

This \textquotedblleft anomalous behavior\textquotedblright\ is typical in
the cases of power-law distributions with divergent mathematical
expectation: $P(n)\simeq n^{-1-\alpha }$ with $0<\alpha <1$. For exponents
in the range $\alpha >1$ the sample mean shows no systematic shift and
fluctuates around the population mean. Specifically \cite{EliSok}: In the
range $1<\alpha <2$ the fluctuations are L\'{e}vy-distributed, and of the
order $O(m^{1/\alpha -1})$. And, in the range $\alpha >2$ these fluctuations
are Normally distributed, and of the order $O(1/\sqrt{m})$.

We considered the problem of sampling from a naturally-truncated power-law
distribution, and the problem of matching two populations with different
naturally-truncated power-law distributions sharing the same population
mean. We have shown that the sample means -- in case of sample sizes which
are considerably smaller than the population size -- fluctuate strongly and
display systematic deviations from the population mean. Since the dependence
of this systematic deviation on the number of sampled elements is known,
this can be used to obtain a rough estimation of the population mean.

\bigskip

I.S. acknowledges the financial support by DFG within the SFB 555
collaboration project. The authors are grateful to J. Klafter for
useful discussions.

\end{document}